\documentclass[aip,apl,reprint]{revtex4-1}
\usepackage{graphicx}% Include figure files
\usepackage{bm}% bold math
\usepackage{color}% bold math
\usepackage[latin1]{inputenc} 

\begin{document}
\title{Quantum dot-cavity strong-coupling regime measured through coherent reflection spectroscopy in a very high-Q micropillar}

\author{Vivien Loo}
\affiliation{Laboratoire de Photonique et Nanostructures, LPN/CNRS, Route de Nozay, 91460 Marcoussis, France}

\author{Loïc Lanco}
\email[]{loic.lanco@lpn.cnrs.fr}
\affiliation{Laboratoire de Photonique et Nanostructures, LPN/CNRS, Route de Nozay, 91460 Marcoussis, France}
\affiliation{Université Paris Diderot - Paris 7, UFR de Physique, 4 rue Elsa Morante, 75205 Paris CEDEX 13, France}

\author{Aristide Lemaître}
\affiliation{Laboratoire de Photonique et Nanostructures, LPN/CNRS, Route de Nozay, 91460 Marcoussis, France}

\author{Isabelle Sagnes}
\affiliation{Laboratoire de Photonique et Nanostructures, LPN/CNRS, Route de Nozay, 91460 Marcoussis, France}

\author{Olivier Krebs}
\affiliation{Laboratoire de Photonique et Nanostructures, LPN/CNRS, Route de Nozay, 91460 Marcoussis, France}

\author{Paul Voisin}
\affiliation{Laboratoire de Photonique et Nanostructures, LPN/CNRS, Route de Nozay, 91460 Marcoussis, France}

\author{Pascale Senellart}
\affiliation{Laboratoire de Photonique et Nanostructures, LPN/CNRS, Route de Nozay, 91460 Marcoussis, France}

\date{\today}

\begin{abstract}
We report on the coherent reflection spectroscopy of a high-quality factor micropillar, in the strong coupling regime with a single InGaAs annealed quantum dot. The absolute reflectivity measurement is used to study the characteristics of our device at low and high excitation power. The strong coupling is obtained with a $g=16\ \mathrm{\mu eV}$ coupling strength in a $7.3 \ \mathrm{\mu m}$ diameter micropillar, with a cavity spectral width $\kappa=20.5 \ \mathrm{\mu eV}$ (Q=65 000). The factor of merit of the strong-coupling regime, $4g/\kappa=3$, is the current state-of-the-art for a quantum dot-micropillar system.

\end{abstract}

\maketitle

High quality factor (Q) optical microcavities embedding semiconductor quantum dots (QDs) are of strong interest for the implementation of cavity quantum electrodynamics in the solid state. The enhancement of the light-matter interaction in the weak coupling regime has recently been used to realize efficient single photon and entangled-photons sources \cite{Strauf2007,Dousse2010}, or for the quantum-dot assisted control of photon transmission \cite{Faraon2008a}. In parallel, the strong coupling regime \cite{Reithmaier2004, Yoshie2004,Peter2005,Englund2007} has been used to implement controlled phase shifts between two laser beams \cite{Fushman2008} or to engineer non-classical states of light using photon blockade \cite{Faraon2008b}. Among the possible cavity geometries, micropillar cavities are of particular interest as the fundamental cavity mode can be coupled to and from the outside with a very high coupling efficiency \cite{Rakher2009}. Moreover, they offer interesting perspectives for the implementation of quantum information protocols using charged quantum dots \cite{Hu2008,Hu2009}.

In this letter, we report on the coherent reflection spectroscopy of a large diameter high-Q micropillar, in a pronounced strong-coupling regime with a single QD transition of large oscillator strength. The characteristics of the system, in particular the QD-cavity coupling strength $g$ and the cavity decay rate $\kappa$, are derived thanks to the very good agreement between the data and a simple theoretical model. The factor of merit of the strong-coupling regime, $4g/\kappa=3$, is the state of the art for a quantum dot embedded in a micropillar cavity \cite{Kistner2010,Reitzenstein2009,Reitzenstein2010}. Using a large diameter micropillar ($7.3 \  \mathrm{\mu m}$ diameter) allows to minimize the side leakage rate, which is mandatory to obtain a high fidelity in several quantum information processes \cite{Hu2008,Hu2009}.

Our sample is grown by molecular beam epitaxy on a GaAs substrate. The bottom (top) distributed Bragg reflector consists in 36 pairs (32 pairs) of alternating quarter-wavelength thick layers of $\mathrm{Ga}_{0.9}\mathrm{Al}_{0.1}\mathrm{As}$ and $\mathrm{Ga}_{0.05}\mathrm{Al}_{0.95}\mathrm{As}$. The GaAs $\lambda$ cavity embeds a single layer of low-density self-assembled In(Ga)As QDs. In order to achieve a large oscillator strength the quantum dots were produced using the dot-in-the-well technique \cite{Nishi1999}. We deposited 1.7 monolayers of InAs on GaAs, stopped the growth for 30 seconds to promote In adatom diffusion towards the quantum dots, and deposited a 5 nm thick In$_{0.15}$Ga$_{0.85}$As layer to further increase the QD size; this led to a QD emission energy around 0.85 eV at room temperature. The sample was then annealed for 30 s at 910°C to induce Ga-In interdiffusion, resulting in a blue-shift of the excitonic transitions and an increased oscillator strength \cite{Langbein2004}. The $7.3 \  \mathrm{\mu m}$ diameter micropillar has been processed through electronic beam lithography and inductively coupled plasma etching \cite{Dousse2009}.

\begin{figure}[ht!]
\includegraphics[width=8.2cm]{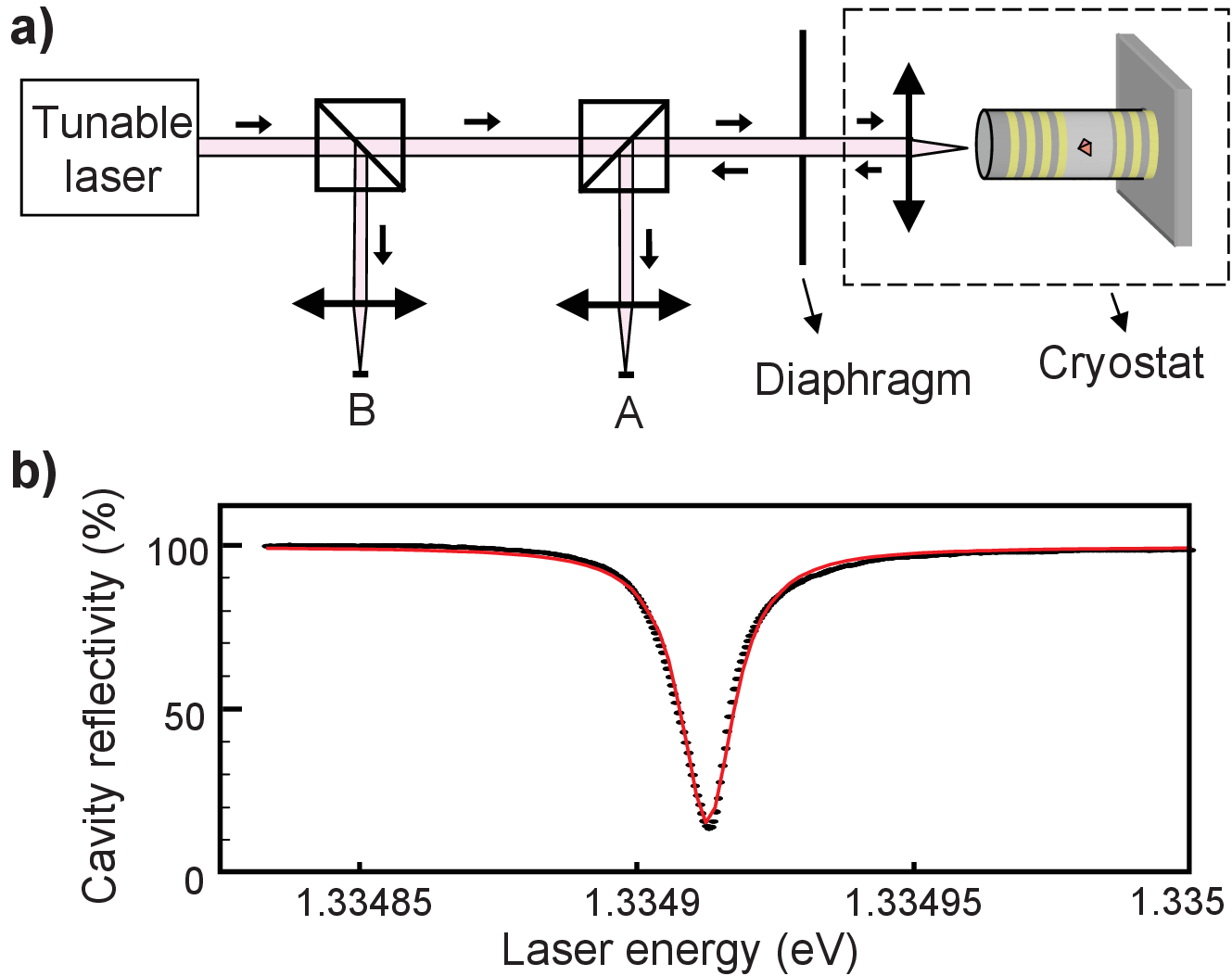}
\caption{\textbf{a)}  Simplified scheme of the experimental setup. \textbf{b)} Cavity reflectivity spectrum measured at $30 \ \mathrm{\mu W}$ excitation power (dots: raw data of the ratio $A/B$ normalized to unity out of resonance, solid line: lorentzian fit).}
\end{figure}

A simplified scheme of the experimental setup is shown in Fig. 1a: a continuous-wave linearly-polarized tunable laser, with a less than $1\ \mathrm{MHz}$ linewidth, is focused on the micropillar surface and the reflected signal is directly measured using a silicon photodiode ($A$). The sample and the focusing lens are placed inside a helium vapor cryostat with temperature control, together with three nanopositioners allowing for a precise adjustment of the beam position. A diaphragm is placed in front of the cryostat to control the beam diameter and thus optimize the mode-matching. Finally, a reference photodiode ($B$) is used to characterize the incident beam power; both photodiodes are linked to current preamplifiers and lock-in amplifiers, and the reflectivity signal is directly measured by the ratio $A/B$. The residual Fabry-Perot interferences, due to the cryostat windows and to the beamsplitters' facets, have been minimized and lead to a less than 2$\%$ modulation on the $A/B$ signal. 

Figure 1b shows the reflectivity spectrum of our micropillar in the case of a high-power excitation ($30 \  \mathrm{\mu}$W). As we will use the $\hbar=1$ convention in the following, the frequencies and decay rates are expressed in electron-volts. The spectrum has a lorentzian shape corresponding to an empty cavity, as all the QD transitions in resonance or near-resonance with the mode are saturated. The fitted full-width at half-amplitude is the cavity decay rate at high-power, $\kappa_{HP}=11.6 \pm 0.05 \ \mathrm{\mu} eV$. This corresponds to a  quality factor at high-power $Q_{HP}=\omega_0/\kappa_{HP}=115000$, where $\omega_0=1.33491 eV$ is the mode resonance frequency. The decay rate $\kappa_{HP}$ is the sum of a mirror-induced decay rate $\kappa_0$ (corresponding intrinsic quality factor $Q_0$) and a loss-induced decay rate $\kappa_{loss}$ which could arise from residual absorption or diffusion through the micropillar leaky modes (side leakage).

The low minimal reflectivity $R_{min}=13\%$ indicates a satisfying value for the input coupling efficiency $\eta_{in}$ (which is a measure of the mode-matching between the incident beam and the cavity mode) but also for the loss rate $\kappa_{loss}$. Indeed, even for a perfect mode matching ($\eta_{in}=1$) the minimal reflectivity is given by $R_{min}=(1-\frac{Q_{HP}}{Q_0})^2=\frac{\kappa_{loss}^2}{\kappa_{HP}^2}$ \cite{Auffeves2007}. This allows us to deduce an  upper bound on the loss rate $\kappa_{loss}/\kappa_{HP}<0.36$, which is compatible with high fidelity quantum information processing \cite{Hu2008, Hu2009}. Similarly, considering the extreme case of a negligible loss-rate $\kappa_{loss}<<\kappa_0$, we find a lower bound for the input coupling efficiency  $\eta_{in}>87\%$.

The coherent reflection spectrocopy of our micropillar at low excitation power (30nW) is presented in Fig. 2, for increasing nominal temperatures between 30.5K and 32K. The reflectivity spectrum shows a clear signature of a QD-cavity resonance in the strong-coupling regime. The double resonance shape arises from the energy splitting between the two mixed exciton-photon eigenstates, when the light-matter interaction is strong enough to allow for several emission-absorption-reemission cycles of the photon during its lifetime in the cavity \cite{Englund2007}. The QD transition becomes visible only near resonance, when the light-matter mixing begins to distort the cavity reflectivity spectrum. At full resonance (near 31.35K) the two mixed states have equal exciton and photon parts, leading to a symetric double-lorentzian shape. When the temperature increases above 31.35K the mixed states separate again, leading to an asymetric reflectivity spectrum and to the progressive disappearance of the QD-induced distortion of the cavity reflectivity.

\begin{figure}[ht!]
\includegraphics[width=8.2cm]{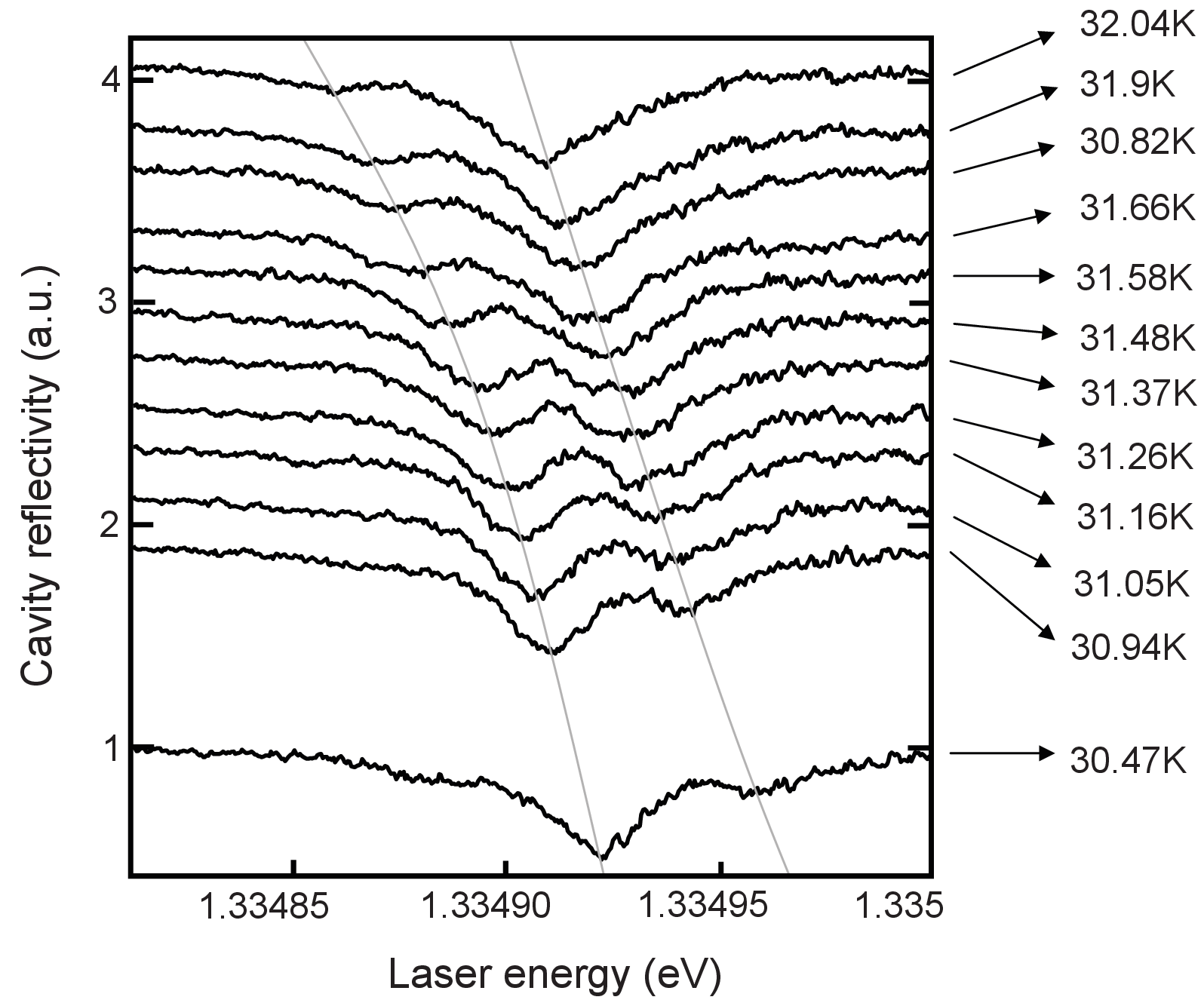}
\caption{Coherent reflection spectroscopy of the QD-micropillar system at $30 \ \mathrm{nW}$ excitation power (raw data), measured for various values of the pillar temperature. Each spectrum is shifted by an amount proportionnal to the corresponding temperature. The thin gray lines are a guide to the eye.}
\end{figure}

Following Englund \emph{et al.} \cite{Englund2007}, we define $\omega_d$ (resp. $\omega_c$) the unperturbed exciton frequency (resp. unperturbed cavity mode frequency), and $\omega$ the incident laser frequency. The quantum dot-cavity interaction is described by its coupling strength $g$, and $\kappa$ is the decay rate of the cavity mode at low excitation power. Our definition of $\kappa$ differs by a factor 2 from that of Englund \emph{et al.}, \cite{Englund2007}, $\kappa$ being the full-width at half-amplitude of the empty-cavity spectrum. Following Auffeves \emph{et al} \cite{Auffeves2007}, we also define $\gamma$ as the sum of two terms, $\gamma=\gamma_{QD}+2\gamma^*$, where $\gamma_{QD}$ is the quantum dot desexcitation rate in all channels other than emission in the cavity mode, and $\gamma^*$ the excitonic dephasing rate. With these notations the reflected signal $R$ from the described micropillar can be derived \cite{Auffeves2007,Waks2006}:
\begin{equation}  \label{eq_reflectivity}
R=\left|1-\sqrt{\eta}\frac{\frac{\kappa}{2}}{\frac{\kappa}{2}+i(\omega-\omega_c)+\frac{g^2}{(\omega_d-\omega)+\gamma}}  \right|^2,
\end{equation}
where $\eta$ accounts for the efficiency of coupling to and collecting from the cavity mode. Basically, the value of $g$ controls the splitting between the two local reflectivity minima, while $\kappa$ is related to their spectral width. Interestingly $\gamma$ does not represent a spectral width directly measurable on the spectrum, but instead governs the contrast of the splitting, that is, the relative height of the reflectivity peak between the two local reflectivity minima.

Figure 3a shows a fit of the reflectivity spectrum at resonance, using Eq. (\ref{eq_reflectivity}), with all fit parameters left free: we find $g=16.1\pm1\  \mathrm{\mu eV}$, $\kappa=20.5\pm4 \  \mathrm{\mu eV}$, $\gamma=17.2\pm2 \ \mathrm{\mu eV}$ and $\eta=43\pm2 \%$. $\omega_c$ and $\omega_d$ are found approximately equal with a less than $1\ \mathrm{\mu eV}$ discrepancy, showing that this spectrum is taken almost at perfect resonance. The value of $\kappa$ corresponds to a quality factor $Q=65 000\pm10000$ at low-power, lower than $Q_{HP}$. This arises from a higher value of the loss-induced decay rate, due to the interaction with the quantum dot layer which is not saturated at low-power. This increase of $\kappa_{loss}$ in the low-power case arises from the sum of small absorption and diffusion processes from numerous quantum dot transitions, which are not spectrally matched but yet in the spectral vicinity of the mode resonance; indeed the spectral density is of the order of 20 quantum dots per meV due to the large micropillar diameter. Considering that $\gamma_{QD}$ is of the order of a few $ \mathrm{\mu eV}$, we also find that $\gamma$ is dominated  by the excitonic dephasing rate $\gamma^*$. This is probably another consequence of the interaction with the non-resonant quantum-dot transitions, as also happens when using non-resonant excitation in photoluminescence experiments \cite{Winger2009}. Working with a quantum dot layer of lower density, especially in the spectral vicinity of the cavity resonance, would probably allow for a large improvement of the low-power quality factor.

\begin{figure}[ht!]
\includegraphics[width=8.2cm]{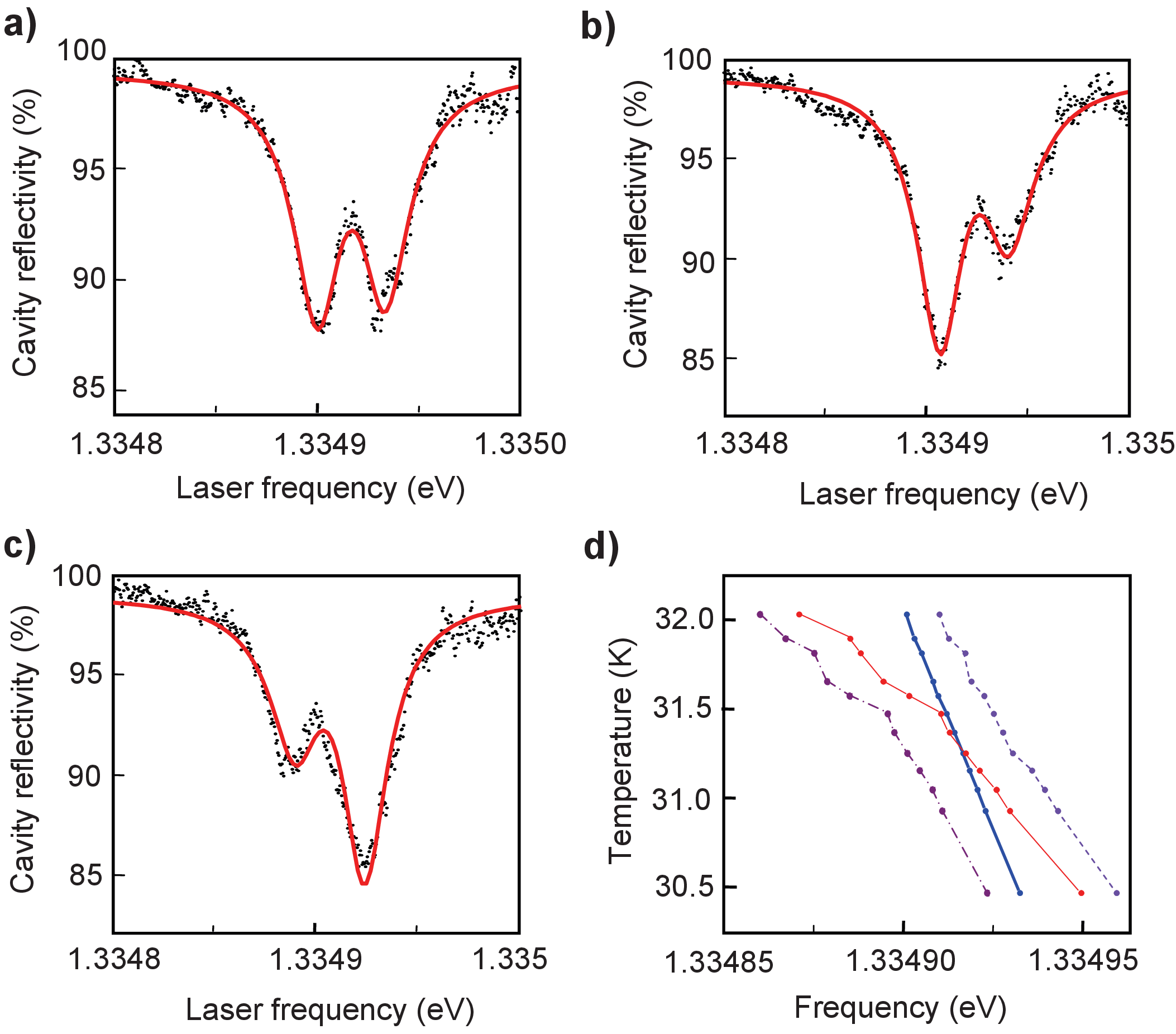}
\caption{\textbf{a)},\textbf{b)} and \textbf{c)}: Normalized cavity reflectivity (dots: experimental data, solid line: fit using Eq. (\ref{eq_reflectivity})) when $\omega_c \approx \omega_d$, $\omega_c <\omega_d$, and $\omega_c > \omega_d$, respectively. \textbf{d)} Unperturbed frequencies $\omega_c$ (bold solid line) and $\omega_d$ (thin solid line), and left and right local reflectivity minima (dashed-dotted line and dashed line, respectively), as a function of temperature.}
\end{figure}

Figure 3b (resp. 3c) shows a fit of the reflectivity spectrum at $31.05$ K (resp. $31.58$ K), where all the parameters are fixed to the above-mentionned values, except for $\omega_c$ and $\omega_d$ which are left free: we find that the shape of the spectra changes with the detuning as predicted by the theory. The unperturbed frequencies $\omega_c$ and $\omega_d$, deduced from fitting the data of Fig. 2, are plotted in Figure 3d as a function of temperature, together with the frequencies of the two local reflectivity minima. Both $\omega_c$ and $\omega_d$ show a quasi-linear dependance with temperature, with $\frac{\rm d \omega_c }{\rm d T}= 20\pm2  \  \mathrm{\mu eV/K} $  and $\frac{\rm d \omega_d }{\rm d T}= 50\pm5 \  \mathrm{\mu eV/K} $, compatible with photoluminescence experiments realized on similar samples \cite{Dousse2008}. In contrast, the two local reflectivity minima show an anticrossing with a splitting of $29.5\pm1  \  \mathrm{\mu eV}$ at resonance, when $\omega_c\approx\omega_d$. Contrary to a $\mathrm{\mu PL}$ experiment, this anticrossing of the local minima is not a feature characteristic of the strong-coupling regime. Indeed, as shown theoretically by Auffeves \emph{et al.} \cite{Auffeves2007} and experimentally by Rakher \emph{et al.} \cite{Rakher2009}, in the weak-coupling regime the reflectivity spectrum is also described by Eq. (\ref{eq_reflectivity}), with two reflectivity gaps on each side of the excitonic peak at $\omega=\omega_{d}$. With our notations, in order to be in the strong coupling regime the quantity $4g$ must be stronger than both $\kappa$ and $\gamma$ \cite{Reithmaier2004}. These measurements thus constitute a clear evidence of the strong-coupling regime, as the main factor of merit $\frac{4g}{\kappa}$ is equal to $3.1\pm0.8$, which is the current state-of-the-art for a quantum dot-micropillar system \cite{Reitzenstein2009,Kistner2010,Reitzenstein2010}. We also obtain $\frac{4g}{\gamma}=3.7\pm0.7$.

Assuming that our quantum dot is located at the center of the cavity mode, we can also deduce its oscillator strength $f$ using :
\begin{equation}
g=\frac{(\pi e^2 f)^{1/2}}{(4 \pi \epsilon m_0 V_{eff})^{1/2}}
\end{equation}
where $\epsilon$ is the dielectric constant of the cavity material, $m_0$ the free electron mass, and $V_{eff}=5.5\ \mathrm{\mu m}^3$ the effective mode volume\cite{Reithmaier2004}. We obtain a $f=50$ oscillator strength, which is five times higher than the typical value for a regular self-assembled InAs/GaAs quantum dot emitting at $1.3\ \mathrm{eV}$ \cite{Warburton1997}. This is a confirmation that rapid thermal annealing, together with the dot-in-the-well epitaxy technique, significantly increases the oscillator strength of In(Ga)As/GaAs quantum dots \cite{Langbein2004}.

In conclusion, we have demonstrated a pronounced strong-coupling regime between a semiconductor quantum dot and a high-Q micropillar cavity, using coherent reflection spectroscopy. The absolute measurement of the reflectivity spectrum allowed us to evidence that these devices can be excited with a very high input coupling ($\eta_{in}>87\%$), as required for the implementation of giant optical nonlinearity at the scale of a few photons \cite{Auffeves2007}. Added to the possible deterministic coupling between quantum dots and micropillars using in-situ lithography \cite{Dousse2008,Dousse2009}, the high value of the factor of merit $\frac{4g}{\kappa}=3$ shows that these devices are very promising for quantum information processes such as photon entanglement \cite{Hu2008,Hu2009} and generation of non-classical states of light \cite{Faraon2008b}. This work is partially supported by the ANR project N°ANR-JCJC-09-0136.

\end{document}